\documentclass[aps,pra,superscriptaddress,twocolumn,longbibliography,amsmath,amssymb]{revtex4-2}

\usepackage{graphicx,cancel}
\usepackage{setspace,color,appendix,physics,wasysym,xcolor,xparse}
\usepackage{appendix}
\usepackage[english]{babel}
\usepackage{amsmath,amssymb,amsfonts,mathtools}
\usepackage[colorlinks = true, urlcolor = black, citecolor = blue, linkcolor = red]{hyperref}

\newcommand{\identity}{1\kern-0.25em\text{l}}

\begin{document}

\title{Continuous Variable Quantum Key Distribution with Single Quadrature Measurement at Arbitrary Reference Frame}
\author{Vinod N. Rao}
\email{vinod.rao@york.ac.uk}
\affiliation{School of Physics, Engineering \& Technology and York Centre for Quantum Technologies, University of York, YO10 5FT York, U.K.}
\affiliation{Quantum Communications Hub, University of York, U.K.}
\author{Emma Tien Hwai Medlock}
\affiliation{School of Physics, Engineering \& Technology and York Centre for Quantum Technologies, University of York, YO10 5FT York, U.K.}
\author{Timothy Spiller}
\affiliation{School of Physics, Engineering \& Technology and York Centre for Quantum Technologies, University of York, YO10 5FT York, U.K.}
\affiliation{Quantum Communications Hub, University of York, U.K.}
\author{Rupesh Kumar}
\email{rupesh.kumar@york.ac.uk}
\affiliation{School of Physics, Engineering \& Technology and York Centre for Quantum Technologies, University of York, YO10 5FT York, U.K.}
\affiliation{Quantum Communications Hub, University of York, U.K.}
 
\begin{abstract}
We propose a simplified measurement scheme for a Gaussian modulated coherent state (GMCS) protocol for continuous variable quantum key distribution (CV-QKD), utilizing homodyne detection without quadrature switching. The reference frame of measurement is taken to be at an arbitrary angle, however, reconciliation converges the proposed scheme to GMCS with switching quadrature protocol. The arbitrary frame of measurement could also include the unknown random thermal drift within Bob's optical measurement setup. We found this scheme is advantageous for practical free-space and fibre-based GMCS protocol based CV-QKD systems as it does not require a phase modulator for random measurement selection quadrature at Bob.
\end{abstract}

\maketitle

\section{Introduction \label{sec:intro}}

Quantum key distribution (QKD) enables two parties to share a secret key, in the presence of an eavesdropper \cite{bennett1984quantum}. The advantage of QKD over the classical counterpart, is that the latter relies on the computational hardness of a problem for security. Whereas, in QKD, information-theoretic security is provided, which is proven by the laws of physics \cite{gisin2002quantum, pirandola2020advances}. The security of QKD relies on no-go theorems, such as, the no-cloning theorem \cite{wootters1982single}, impossibility in perfectly distinguishing non-orthogonal states \cite{ivanovic1987differentiate}, monogamy of entanglement \cite{horodecki2009quantum}, the uncertainty principle among two non-commuting observable \cite{koashi2006unconditional} and so on. The earlier proposals \cite{bennett1984quantum, ekert1991quantum, bennett1992quantum} and demonstrations \cite{marand1995quantum, hughes2000quantum, stucki2002quantum, wang2005experimental} for QKD were based on discrete variables (DV) using weak coherent states, based on polarization or phase qubits and single photon detectors. Recently, newer protocols were developed and demonstrated for longer transmission distances \cite{lo2012measurement, lucamarini2018overcoming, wang2022twin}.

Continuous variable (CV) QKD protocols \cite{ralph1999continuous, hillery2000quantum, reid2000quantum} utilize amplitude and phase modulation of coherent states to encode the secure key information and shot noise limited detection for decoding. Squeezed as well as thermal states based protocols have also been proposed for implementing CV-QKD protocols \cite{ralph1999continuous}. Among the various CV-QKD protocols, the Gaussian Modulated Coherent State (GMCS) protocol \cite{grosshans2002continuous, braunstein2005quantum}, has been thoroughly studied for its security and demonstrated over fibre as well as free-space channels. The amplitude and phase of the coherent state are modulated such that the field quadratures follow the Gaussian probabilistic distribution.

There are two versions of the GMCS protocol based on the measurement of the quadratures. In switching quadrature protocol (`GG02')\cite{grosshans2002continuous}, Bob randomly selects the quadrature for the measurement using a homodyne detector. In the non-switching version of the GG02 protocol \cite{weedbrook2004quantum}, Bob measures both quadratures with a Heterodyne detector. The non-switching protocol has higher information decoding capacity for shorter distances \cite{andersen2010continuous, soh2015self, zhang2019continuous}, while the switching protocol is more suitable for long distances CV-QKD demonstrations \cite{jouguet2013experimental, huang2016long, zhang2020long}. At the detectors, the relative phase of $0^\circ$ or $90^\circ$, between the CV-QKD signal with a strong reference signal - referred to as the Local Oscillator (LO), sets the reference frame for the quadrature measurement.

There are also two variants for CV-QKD protocol implementation. In the Transmitting Local Oscillator (TLO) scheme, the LO is transmitted along with the CV-QKD signal \cite{grosshans2002continuous}. Since both the signal and the LO pass through the same channel, the phase drift from the channel is negligible. However, with the Local Local Oscillator (LLO) scheme, only the reference pulse is sent by Alice to Bob and the LO is locally generated inside Bob's station \cite{soh2015self}. 

In theoretical analysis, the reference frames of state preparation at Alice and that of the state measurement at Bob are assumed to be fixed and known to Eve. To avoid Eve replacing the coherent states with squeezed states, either a random selection of the quadrature or measuring both of them at the same time is necessary. However, in practice, the frame of measurement is continuously rotated at an arbitrary rate. Measurement of quadrature along the arbitrary frame of reference has been already proposed in \cite{soh2015self}, however this is a quadrature switching version with the homodyne detector. We propose the arbitrary frame of measurement scheme without switching quadratures and using a homodyne detector. We consider the measurement frame of reference to be dynamic during the protocol run. The initial angle, the rate of rotation and the direction of the rotation of the frame of measurement are taken to be random as well. We also assume that over the protocol run, the frame of measurement could be rotated to all the possible angles. The phase drift can collectively be caused by - Alice's preparation setup - $\theta_{\text{Alice}}$, transmission channel - $\theta_{\text{ch}}$ and Bob's measurement setup - $\theta_{\text{Bob}}$. We note here that the current work addresses the phase drift that happens at Bob's measurement setup inside his station, and works for both transmitted/local local oscillator schemes.

In this paper, we aim to validate that the GG02 protocol can be performed using homodyne detection and without actively switching the quadratures as originally proposed and demonstrated over the last two decades. We prove that our scheme retains the characteristics of the GG02 protocol and, thus does not create any new security loopholes. Specifically, the aspect of security that is considered here is for the collective attack by the eavesdropper Eve and we show that the protocol is secure in the asymptotic limit. We also show an optimal intercept and resend attack by Eve, and highlight that it is different from the one in GG02-like protocols. Additionally, our proposed scheme works not only with a TLO setup, but with a LLO setup as well. In the LLO variant of the proposed protocol, all the quantum parts would remain same as in a standard LLO protocol. The only difference would be that Bob sets his LO to a random quadrature in the phase space and then correlates to the reference pulse sent by Alice.

The paper is structured as follows. The Sec. \ref{sec:gmcv} provides the outline of Gaussian modulated CV-QKD protocol GG02. In Sec. \ref{sec:rota}, we describe the arbitrary rotation of the frame of reference in Bob's measurement and prove its equivalence to GG02. Finally, discussion and conclusion of our results is in Sec. \ref{sec:disc}.

\section{Gaussian Modulated Coherent State Protocol \label{sec:gmcv}}
A brief introduction to conventional GMCS protocols is given below. These protocols utilise coherent states, $\ket{\alpha} = \ket{\hat{q}_{\text{A}}+i \hat{p}_{\text{A}}}$, as the information carriers, wherein the state quadratures, $\hat{q}_{\text{A}}$ and $\hat{p}_{\text{A}}$, are randomly modulated by Alice such that the quadratures follow Gaussian distribution, $\mathcal{N}(0,\mathbb{V}_{\text{A}})$, of variance $\mathbb{V}_{\text{A}}$ and mean at zero. Along with the signal states, Alice also sends the LO pulses and thus involves them of being suitably multiplexed by Alice and demultiplexed by Bob. The signal states undergo attenuation and noise is added to it during the transmission from Alice to Bob. At Bob, the quadratures of the modulated coherent states, $\hat{q}_{\text{B}}$ and $\hat{p}_{\text{B}}$, are measured with a shot noise (vacuum noise) limited homodyne (switching protocol) or heterodyne detector (non-switching protocol). The variance of the measurement at Bob can be written as $\mathbb{V}_{\text{B}} = T * \mathbb{V}_{\text{A}} + \xi_t$, in which $T$ is the transmittance of the channel and $\xi_t$ is the total noise variance added to the quadratures.

During the classical post-processing, Bob publicly announces the chosen quadrature for measurement and Alice chooses the respective variance. Additionally, Bob also announces the variance $\mathbb{V}_{\text{B}}$ of a fraction of transmitted states, to Alice for quantifying error in the channel. The remaining data is used for secure key generation after error correction and privacy amplification. In the switching version of the protocol, Bob has to disclose the quadrature he measures (not the measurement outcomes) as well and that forms the sifting. The noise variance term $\xi_t$ comprises of the shot noise variance, electronic noise variance of the detector, noise from eavesdropping among others. One can assume that the channel transmittance $T$, along with the excess noise, is controlled by Eve. 

The feature that prevents the possibility of CV-QKD with fixed quadrature measurement is the following. By knowing the frame of Bob's measurement, Eve can launch the Intercept and Resend (IR) attack, during which she can resend a suitable squeezed state along the frame of quadrature measurement and hides her presence. Given Bob's selection of quadrature for the measurement (among $\hat{q}_{\text{B}}$ and $\hat{p}_{\text{B}}$) is unknown to Eve, she has $50\%$ chance to hide her attack with squeezed states. To prevent Eve from using a squeezed state, one has to either switch between the quadratures randomly \cite{grosshans2002continuous} or measure both quadratures \cite{weedbrook2004quantum}.

Therefore, both versions of the CV-QKD protocol offers security against individual attacks by Eve. In case of collective attack such as Eve using an entangled cloner, she entangles her own n-mode squeezed states with the state sent by Alice. In the switching version of the protocol, she waits for Bob to announce his choice of quadrature to and measures her n-mode state (stored in the quantum memory). In no-switching version, Eve does not need to wait for Bob's announcement. However, as in the case of individual attack, the action of entangling $\ket{\alpha}$ with her own n-mode squeezed state increases the excess noise \cite{lupo2020towards}.

For the switching protocol, a specific quadrature for measurement at Bob is chosen at random, by suitably selecting the phase of the LO. The relative phase of LO at Bob with respect to the initial phase during the coherent state preparation defines the chosen quadrature for measurement, which are conventionally taken to be $0^\circ$ for $\hat{q}_{\text{A}}$ quadrature and $90^\circ$ for $\hat{p}_{\text{A}}$ quadrature. However, in practice, the relative phase between signal and LO inside Bob's station is continuously drifting due to thermal fluctuations and relative frequency drift. Primarily, the fluctuations could be due to the different paths that the signal and LO take after de-multiplexing within Bob's measurement setup. Additionally, in the case of the LLO scheme, it can happen due to the frequency drift of the laser at Bob's station. Limiting our arguments in the rest of the manuscript to phase drift inside Bob's station $\theta_{\text{Bob}}$ alone, we attribute this relative phase drift to the signal. If the phase drift is relatively slow compared to the QKD repetition rate, which is a requirement for our scheme to work, this can be associated to phase noise \cite{soh2015self}. However, if the repetition rate is slower than that of the phase drift then it requires heterodyne detection for phase estimation. Also, the key rate would significantly drop, as the correlation between Alice and Bob would reduce. This phase drift is unknown to Bob, but he could easily estimate it after measurement. In principle, Eve could probe this phase drift with a very high intense beam but there are countermeasures to prevent such attacks \cite{gisin2006trojan}. Also, high-intensity probe beams saturate the measurements and can be detected as well \cite{qin2016quantum}. The phase drift creates an additional post-processing routine, as either Bob or Alice has to rotate their frame of reference to compensate for the drift and match reference frames with each other. Therefore, even assuming that Eve controls the drift in the channel, she would still have no knowledge of the phase drift inside Bob's (or Alice's) station. There are proposals to use machine learning algorithms for estimating the phase drift in the channel \cite{xing2022phase}, however this is beyond the scope of the current work.

In the case where Bob actively compensates for phase drift in real-time, a fixed choice of phase enables Eve to launch IR attack as the frame of measurement at Bob matches with that of state preparation at Alice. But, in the case where Alice compensates her quadrature data for the phase drift during the post-processing and thereby matches with Bob's frame of measurement, a single (fixed) quadrature measurement can be implemented without switching the quadrature. In the next section, we provide evidence that no switching GMCS protocol with single quadrature measurement at an arbitrary choice of $\theta_{\text{Bob}}$ is equivalent to GG02 protocol.

\begin{figure}[h]
\includegraphics[width=\linewidth]{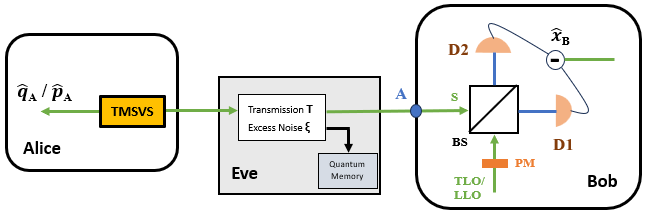}
\caption{CV-QKD with the homodyne detection setting. Alice prepares two mode squeezed vacuum states (TMSVS) and sends one of the modes to Bob. Bob's setup interferes the signal (S) with the Local Oscillator (LO) pulses using a beam splitter (BS) and performs homodyne measurements using photodiodes (D1 \& D2). There could be a relative phase drift between signal and LO inside Bob's station (from the blue marker-`A' to the BS) represented by $\theta_{\text{Bob}}$ (see text below), that is independent from the phase drift in the channel and Alice's station. Contrary to standard homodyne measurement, we do not use a phase modulator (PM) in our scheme. $\hat{x}_{\text{B}}$ is the quadrature output of the homodyne along the measurement angle $\theta_{\text{Bob}}$. The eavesdropping by Eve is characterized by reduced transmittivity $T$, and excess noise from the channel $\xi$.}
\label{fig:cvqkd}
\end{figure}

\section{Arbitrary rotation in measurement frame of reference \label{sec:rota}}

Continuing with the discussions of GMCS protocol given in the previous section, we follow the prepare and measure (PM) based description of the protocol. The unconditional security of the PM-based protocols can be verified by mapping it to an equivalent of the entanglement based (EB) protocol. In the EB scheme, Alice prepares two-mode squeezed vacuum states (TMSVS), keeps one of the modes with herself, and sends the other to Bob as in Fig. \ref{fig:cvqkd}. The covariance matrix formalism helps to find the respective mutual information quantities between Alice and Bob, as well as that of Eve with Alice or Bob. Here, the covariance matrix is obtained by finding the expectation values of the respective quadratures. Since Alice's quadratures are $\hat{q}_{\text{A}}$ and $\hat{p}_{\text{A}}$, we find the first entry in the matrix as, $\langle \hat{q}_{\text{A}}^2 \rangle - \langle \hat{q}_{\text{A}} \rangle^2$ and similarly for the other elements. The covariance matrix of the TMSVS is,
\begin{align}
\Sigma =
\begin{pmatrix}
&V\identity_2 & \sqrt{V^2-1}\sigma_z \\
&\sqrt{V^2-1}\sigma_z &V\identity_2
\end{pmatrix},
\label{eq:ABcov}
\end{align}
where $\identity_2$, $\sigma_z$ are the respective Pauli matrices, $V$ is the variance of the squeezed states. This is for the case wherein channel transmittivity $T=1$ and noise $\xi=0$. The corresponding rows (and columns) here represent the variances of Alice's quadratures, $\hat{q}_{\text{A}}, \hat{p}_{\text{A}}$ of mode-1, Bob's quadratures, $\hat{q}_{\text{B}}$ and $\hat{p}_{\text{B}}$, of mode-2, such that $[\hat{q}_{\text{A}}, \hat{p}_{\text{A}}]=[\hat{q}_{\text{B}}, \hat{p}_{\text{B}}]=i/2$. The respective quadratures can be perfectly correlated during post processing, such that $\hat{q}_{\text{A}}=\hat{q}_{\text{B}}$ and $\hat{p}_{\text{A}} = \hat{p}_{\text{B}}$. Bob measures either both the quadratures of mode-2 simultaneously with a heterodyne detector, or one of them at random with a homodyne detector. We find that the variance, $V(\hat{q}_{\text{B}}) = V(\hat{p}_{\text{B}}) = V$, identical to that of Alice's quadratures.

\begin{figure}[h]
\includegraphics[width=\linewidth]{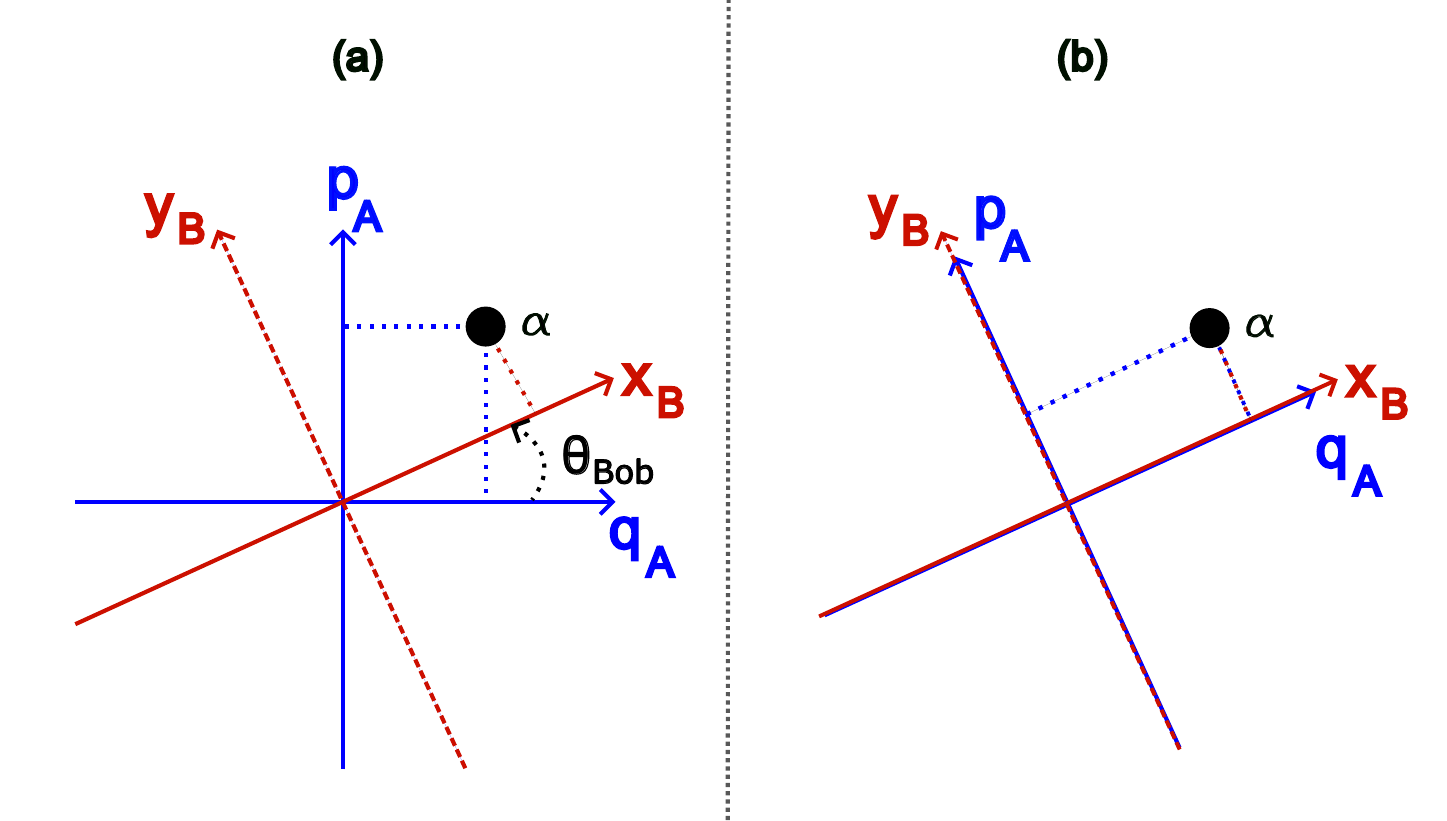}
\caption{(a) Rotated frame of measurement by the receiver Bob. He always measures along the quadrature $\hat{x}_{\text{B}} = \cos{\theta_{\text{Bob}}}(\hat{q}_{\text{B}}) + \sin{\theta_{\text{Bob}}}(\hat{p}_{\text{B}})$, where $\theta_{\text{Bob}}$ is the angle of the rotation w.r.t $\hat{q}_{\text{B}}$ (as well as $\hat{q}_{\text{A}}$). (b) After Bob's announcement of the angle $\theta_{\text{Bob}}$, Alice rotates her frame to coincide with $\hat{x}_{\text{B}}$.}
\label{fig:frame}
\end{figure}

Consider the case, wherein Bob chooses to perform single quadrature measurement of the mode-2 throughout the protocol, at an arbitrary angle of measurement, $\theta$ - thanks to the phase drift, as shown in Fig. \ref{fig:frame}. For convention, we take that he chooses to measure in the $\hat{x}_{\text{B}} = \cos{\theta_{\text{Bob}}}(\hat{q}_{\text{B}}) + \sin{\theta_{\text{Bob}}}(\hat{p}_{\text{B}})$ quadrature. While the other quadrature would be $\hat{y}_{\text{B}}$, with $[\hat{x}_{\text{B}}, \hat{y}_{\text{B}}]=2i$, we shall restrict the case to Bob always choosing quadrature $\hat{x}_{\text{B}}$ for measurement. In other words, Bob only acquires the homodyne detector outcome and refers to it as $x_{\text{B}}$. After Bob's announcement of the angle $\theta_{\text{Bob}}$, Alice rotates her frame to coincide with $\hat{x}_{\text{B}}$. However, there could be a small phase mismatch between Alice and Bob as the phase is drifting continuously at Bob's station, which is addressed in Figs. \ref{fig:key} and \ref{fig:correlation}.

In the following, we consider three different scenarios in which the measurement frame of reference, $\theta_{\text{Bob}}$, is: (a) fixed and known to Eve; (b) fixed and unknown to Eve; and (c) varying and unknown to both Eve and Bob.

\subsection{$\theta_{\text{Bob}}$ is fixed and known to Eve}

Here we consider the reference frame of measurement is rotated by a fixed angle $\theta_{\text{Bob}}$ and Eve has the knowledge about the rotation. She is also aware that Bob will be measuring only one of the quadratures ($\hat{x}_{\text{B}}$) throughout the protocol run. In this scenario, Eve can successfully perform a trivial IR attack without introducing any error in Bob's measurement. Though the variance of the other quadrature ($\hat{y}_{\text{B}}$) increases due to eavesdropping, since it is never measured, the protocol becomes completely insecure. This would be identical to Eve knowing Bob's choice of quadrature for each pulse in the homodyne based conventional CV-QKD protocol. The requirement of $\theta_{\text{Bob}}$ being unknown to Eve makes the single quadrature measurement strategy interesting and secure. We will analyse this case in the following subsection.

\subsection{$\theta_{\text{Bob}}$ is fixed and unknown to Eve}

Here, we consider Bob's reference frame of measurement is rotated by a fixed angle $\theta_{\text{Bob}}$, as in the previous case. Additionally, we take that Eve does not have knowledge of the angle $\theta_{\text{Bob}}$ until the announcement by Bob via public channel, as in the case of GG02 of Eve not knowing the chosen quadrature until announcement. We also assume that there are proper countermeasures implemented to prevent Eve from probing the angle, such as an optical isolator at Bob's input. So we limit our arguments to a trusted device scenario, and this assumption is equivalent to the one in GG02 protocol, of using countermeasures to check for Trojan horse like attacks \cite{pan2020practical}. We find the covariance matrix considering Bob measuring with angle $\theta_{\text{Bob}}$ and choosing only $\hat{x}_{\text{B}}$ all the times. The matrix elements $\Sigma_{11}, \Sigma_{12}, \Sigma_{21}, \Sigma_{22}$ are identical to that in Eq. (\ref{eq:ABcov}). However, other entries are different, in that $\Sigma_{33} = \langle \hat{x}_{\text{B}}^2 \rangle - \langle \hat{x}_{\text{B}} \rangle^2$, and so on. Thus the covariance matrix is found to be,
\begin{align}
\Sigma_{\text{AB}} =
\begin{pmatrix}
&V\identity_2 & \sqrt{V^2-1}P \\
&\sqrt{V^2-1}P^{\text{T}} &V\identity_2+\frac{(V+1)}{2}R
\label{eq:ABcov2}
\end{pmatrix},
\end{align}
as we find that $V(\hat{x}_{\text{B}}) = V(\hat{p}_{\text{B}}) = V$, and where $P=\{\{\cos{\theta_{\text{Bob}}},0\},\{-\sin{\theta_{\text{Bob}}},-1\}\}$ and $R=\{\{0,\sin{\theta_{\text{Bob}}}\},\{\sin{\theta_{\text{Bob}}},0\}\}$. Also, keeping $\theta_{\text{Bob}} = 0$, we get Eq. (\ref{eq:ABcov}).

Since $\theta_{\text{Bob}}$ is unknown to Eve, an IR attack will not be successful, as the interception would increase the variance with a high probability, when Bob measures. Thus the excess noise induced by the attack would lead to termination of the protocol. Here we consider Eve's collective attack via an entangled cloner, wherein she entangles her probe with mode-2 of the TMSVS sent by Alice to Bob. Eve's cloner could be n-mode, and use one of the modes to entangle with mode-2. However, this can be reduced to Eve using a TMSVS and using one of the modes to entangle, due to monogamy of entanglement. Thus we consider Eve's TMSVS and the covariance matrix of her entangling cloner, similar to that of Alice-Bob in Eq. (\ref{eq:ABcov2}), is
\begin{align}
\Sigma_{\text{E}} = 
\begin{pmatrix}
&W\identity_2 & \sqrt{W^2-1}Q \\
&\sqrt{W^2-1}Q^{\text{T}} &W\identity_2+\frac{(W+1)}{2}S
\end{pmatrix},
\label{eq:evetmsvs}
\end{align}
where $\Sigma_{\text{E}} = \Sigma_{\text{E}_1\text{E}_2}$ corresponds to the Eve's TMSVS (with $\text{E}_{1,2}$ representing the respective mode), $Q=\{\{\cos{\phi},0\},\{-\sin{\phi},-1\}\}$ and $S=\{\{0,\sin{\phi}\},\{\sin{\phi},0\}\}$, and $\phi$ being the angle of measurement by Eve. Assuming that Eve replaces the lossy quantum channel with a beam-splitter (BS) of transmission $T$, the symplectic BS \cite{laudenbach2018continuous} acting on the mode-2 and $\text{E}_1$ is represented by,
\begin{align}
\Sigma_{\text{BS}} = 
\begin{pmatrix}
\identity_2 & 0 & 0 & 0 \\
0 & \sqrt{T}\identity_2 &\sqrt{1-T}\identity_2 & 0 \\
0 & -\sqrt{1-T}\identity_2 & \sqrt{T}\identity_2 & 0 \\
0 & 0 & 0 & \identity_2
\end{pmatrix}.
\end{align}

The action of the BS on the joint system of Alice, Bob and Eve is described by,
\begin{equation}
\Sigma^\prime = \Sigma_{\text{BS}} \Sigma_{\text{ABE}} \Sigma_{\text{BS}}^{\text{T}},
\end{equation}
with $\Sigma_{\text{ABE}} = \Sigma_{\text{AB}} \oplus \Sigma_\text{E}$ corresponds to the joint system of Alice, Bob and Eve. 

Using $\Sigma^\prime$ in Eq. (\ref{eq:sigmaprime}), we find the reduced covariance matrix of Alice-Bob from as
\begin{align}
\Sigma^\prime_{\text{AB}} = 
\begin{pmatrix}
V\identity_2 & \mathcal{X}_{+}P \\
\mathcal{X}_{+}P^{\text{T}} & \mathcal{Z}_1
\end{pmatrix},
\label{eq:sigmaprimeab}
\end{align}
and that of Eve is
\begin{align}
\Sigma^\prime_{\text{E}} = 
\begin{pmatrix}
\mathcal{Z}_2 & \mathcal{Y}_{+}Q \\
\mathcal{Y}_{+}Q^{\text{T}} & W\identity_2+\frac{(W+1)}{2}S
\end{pmatrix},
\end{align}
where $\mathcal{X}_{+}, \mathcal{Y}_{+}, \mathcal{Z}_1, \mathcal{Z}_2$ are given in Appendix \ref{app:calc}.

From Eq. (\ref{eq:sigmaprimeab}), the respective mode variances are, $V_{\text{A}}=V$ [$=V(\hat{q}_{\text{A}}) = V(\hat{p}_{\text{A}})$] and $V_{\text{B}}= TV+(1-T)W$ [$=V(\hat{x}_{\text{B}})$]. For $T=1$, we obtain Eq. (\ref{eq:ABcov}). As expected, we find that $V(\hat{q}_{\text{B}})= V(\hat{p}_{\text{B}}) = V(\hat{x}_{\text{B}})$ [$=V(\hat{y}_{\text{B}})$]. Since the rotation in the frame of measurement is a unitary operation on mode-2, the variance does not change. Thus Bob measures as much variance in a rotated frame as he would have for the standard $\hat{q}$ or $\hat{p}$ quadrature. After the announcement of $\theta_{\text{Bob}}$ via public channel by Bob and establishing the correlation with Alice's quadrature data values, they estimate their mutual information as follows. By choosing Eve's variance to be $W=1+\frac{\xi}{1-T}$ \cite{laudenbach2018continuous}, we find $V_{\text{B}}=T(V-1)+1 + \xi$. Therefore, the signal variance at Bob's station is given by, $V_s = T(V-1)$ and the noise variance is $V_n = 1 + \xi$. Thus, we find the mutual information between Alice and Bob to be \cite{weedbrook2012gaussian}
\begin{align}
I(\text{A}:\text{B}) = \frac{1}{2}\log_2\bigg(1+\frac{V_s}{V_n}\bigg) = \frac{1}{2}\log_2\bigg(1+\frac{T(V-1)}{1+\xi}\bigg),
\label{eq:IAB}
\end{align}
where $\xi$ is the excess noise. Here, we note that the mutual information between Alice and Bob is identical to that in the case when Bob performs homodyne measurements along the conventional ($\hat{q}$ or $\hat{p}$) quadratures \cite{laudenbach2018continuous}.

To quantify the information leaked to Eve, Alice and Bob need to estimate $\chi = S_{\text{E}} - S_{\text{E}\mid \text{B}}$, where $S_{(x)} = -\text{Tr}[x \log(x)]$ corresponds to the von Neumann entropy. Given $\rho_{\text{AB}}$ represents the joint state of Alice, Bob and Eve, let $\rho_{\text{AB}}=\text{Tr}_{\text{E}}(\rho_{\text{ABE}})$ and $\rho_{\text{E}}=\text{Tr}_{\text{AB}}(\rho_{\text{ABE}})$. Since the von Neumann entropy depends only on the coefficients of the Schmidt decomposition (which are found to be identical for $\rho_{\text{AB}}$ and $\rho_{\text{E}}$ \cite{weedbrook2012gaussian}), we find that $\chi = S_{\text{E}} - S_{\text{E}\mid \text{B}} = S_{\text{AB}} - S_{\text{A}\mid \text{B}}$. To find $S_{\text{AB}}$, we find the symplectic eigenvalues of the covariance matrix $\Sigma^\prime_{\text{AB}}$. Similarly, we find $S_{\text{A}\mid \text{B}}$ by the symplectic eigenvalues of the covariance matrix $\Sigma^\prime_{\text{A}\mid \text{B}}$, taking the partial measurements on the covariance matrix as given in Eq. (\ref{eq:cova}). The elucidation of the same is given in Appendix \ref{app:mut}.

\begin{figure}[h]
\includegraphics[width=\linewidth]{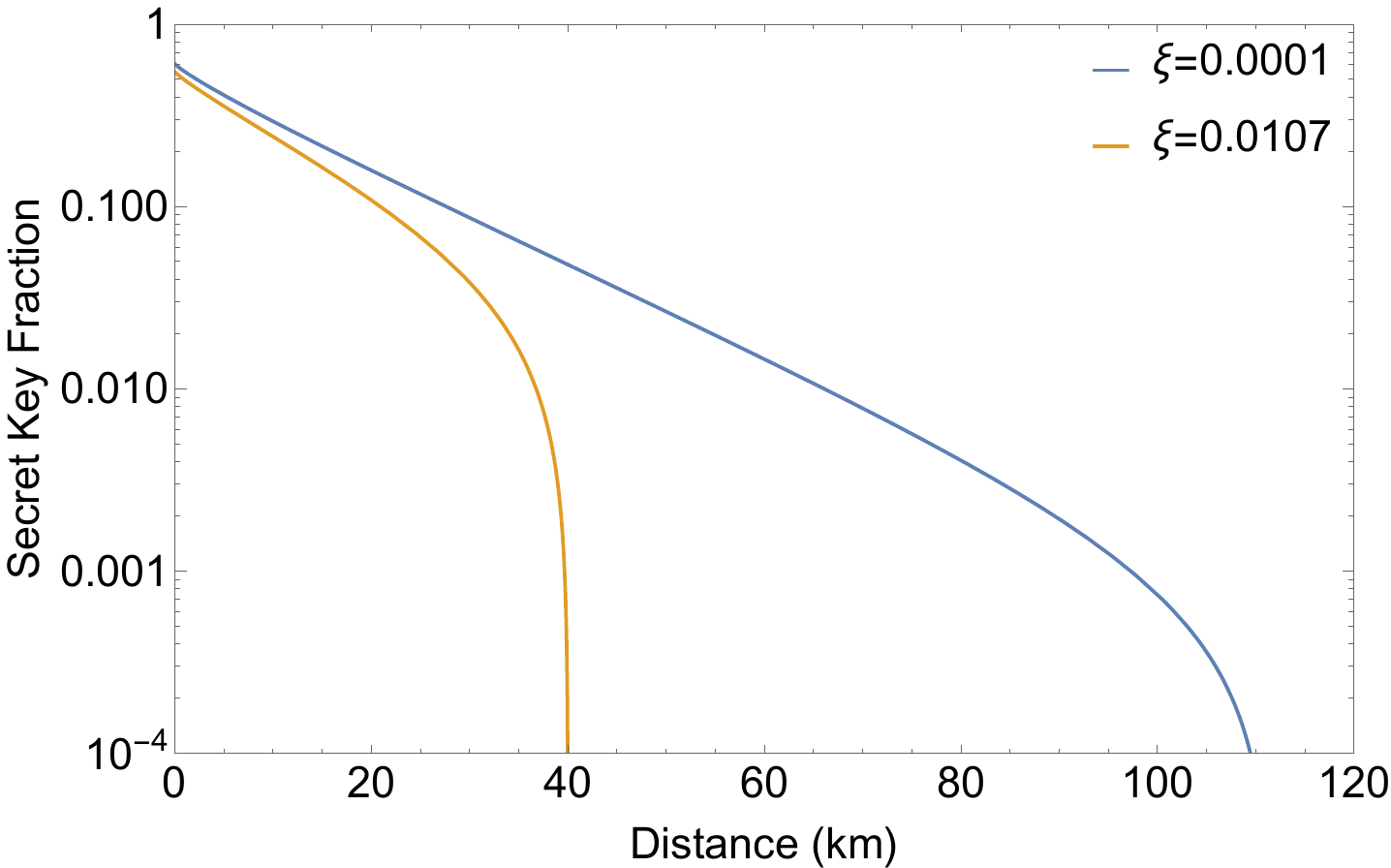}
\caption{Secret key fraction as a function of distance. Here the angle of rotation is chosen to be $\theta_{\text{Bob}} = 90^\circ$. Blue (Orange) line corresponds to the noise $\xi=0.0001$ ($\xi=0.0107$, for a phase drift of $5^\circ$).}
\label{fig:key}
\end{figure}

The secret key fraction is thus found to be,
\begin{align}
r &= \beta ~I(\text{A}:\text{B}) - \chi \nonumber \\
&= \frac{\beta}{2}\log_2\bigg(1+\frac{T(V-1)}{1+\xi}\bigg) - g(\nu_{+}) - g(\nu_{-}) + g(\nu),
\end{align}
where $\nu_{+}, \nu_{-}, \nu$ are the respective symplectic eigenvalues of $S_{\text{AB}}$ and $S_{\text{A}\mid \text{B}}$, given in Eq. (\ref{eq:nupm}) and Eq. (\ref{eq:nu}). The Fig. \ref{fig:key}, corresponds to the secret key fraction after the public announcement of $\theta_{\text{Bob}}$ by Bob to Alice. Here the frame of reference is kept along the $\hat{p}_{\text{B}}$ quadrature. We also include the possible key rate for a potential phase drift of $5^\circ$, that corresponds to an excess noise of $\xi=0.0107$ \cite{tang2020performance}. We note that this phase drift is assumed to be happening within Bob's station. This phase error could be due to inaccurate phase estimation by Bob, which is fundamentally limited by the shot noise with the pattern signal measurements. In TLO scheme, the phase drift is estimated using the pattern signals which are comparatively at higher intensity than the QKD signals and are sent with each block of QKD signals \cite{cosijns2018advanced, tang2021towards, tang2022experimental}. However, the residual phase mismatch is common in any CV-QKD system, independent of the receiver architecture and protocol.

\begin{figure}[h]
\includegraphics[width=\linewidth]{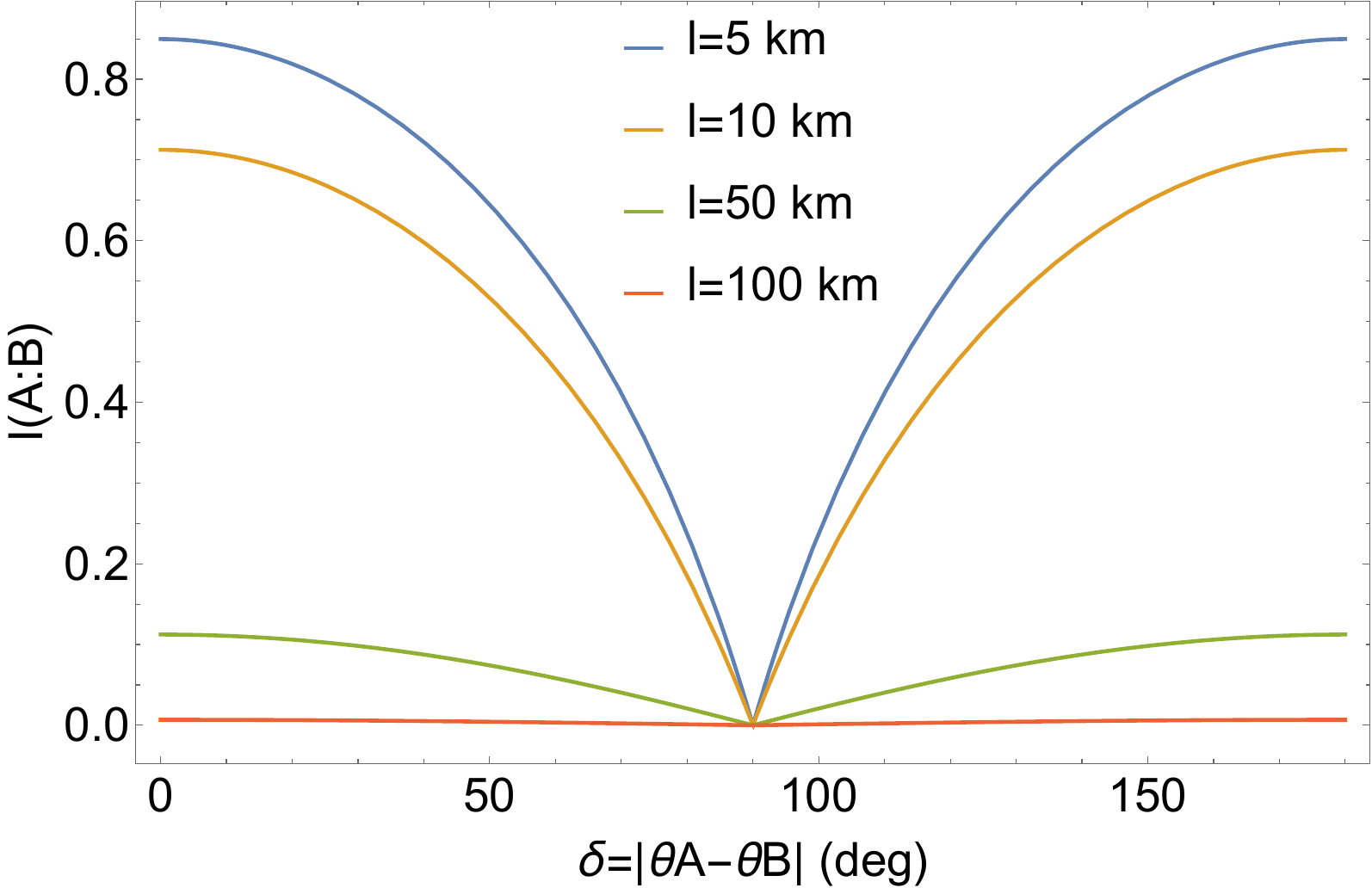}
\caption{Mutual information between Alice and Bob, as a function of the relative angle of the frame of references.}
\label{fig:correlation}
\end{figure}

In Fig. \ref{fig:correlation}, we plot the mutual information between Alice and Bob with the variation in the angle of rotation in their frame of measurements prior to the public announcement of $\theta_{\text{Bob}}$ by Bob. In particular, similar to Bob, Alice also measures the variance with rotating her frame of measurement to $\theta_{\text{A}}$ which is independent of Bob's rotation. The mutual information then is found to be
\begin{equation}
I(\text{A}:\text{B}) = \frac{1}{2}\log_2\bigg(1+\frac{T(V^\prime-1)}{1+\xi}\bigg),
\label{eq:IAB2}
\end{equation}
similar to the earlier case in Eq. (\ref{eq:IAB}), but where $V^\prime = (V_{\text{a}}\cos{\theta_{\text{Bob}}})/\cos{\theta_{\text{A}}}$ is Bob's variance and $V_{\text{a}}$ is Alice's variance along $\theta_{\text{A}}$. We get maximum information for the relative angle of the frame of references $\delta \mid\theta_{\text{Bob}} - \theta_{\text{A}} \mid = 0$, showing perfect coinciding of Alice's and Bob's quadratures after the public announcement. Various distances are considered that also provides the upper bound on the respective mutual information.

\subsection{$\theta_{\text{Bob}}$ is drifting, and unknown to Bob \& Eve}

The practical implementation of CV-QKD experiences random drift in $\theta_{\text{Bob}}$ which is unknown before the measurement but can be monitored by Bob. The aspect of quantifying the phase drift in Bob's measurement is already explored in various works \cite{soh2015self, huang2016long, zhang2020long, liu2021homodyne, jain2022practical}. This rotation is typically slow compared to the CV-QKD signal generation and detection rate. Specifically, the phase drift inside Bob's station could be due to various factors, including thermal effects. If the phase drift is relatively slow and/or the linewidth of the laser is narrow, the phase noise within the QKD data post-processing blocks is minimal \cite{soh2015self}. However, if the phase drift is quick and impulsive, the phase noise plays a role in limiting the achievable transmission distance in CV-QKD \cite{qi2016simultaneous, marie2017self}.

\begin{figure}[h]
\includegraphics[width=\linewidth]{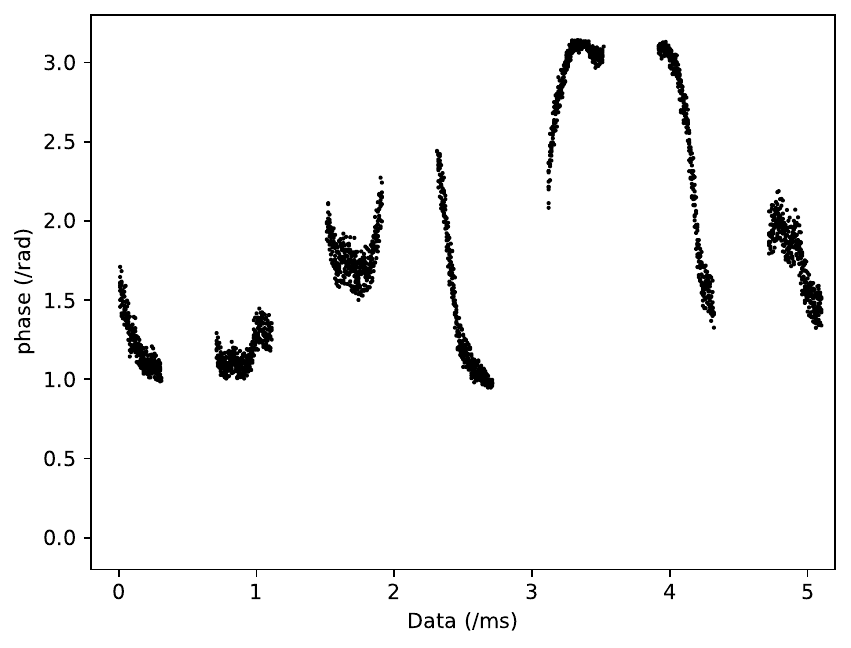}
\caption{Experimentally measured phase drift $\theta_{\text{Bob}}$ inside Bob's station, from the point `A' to BS in Fig. \ref{fig:cvqkd}. The figure shows the randomness in the phase drift fluctuations along with the frequency of phase drift. We have taken the data with an asymmetric Mach-Zehnder interferometer, having 100 ns delay between the two paths - of signal and LO. This in principle emulates the phase drift due to thermal fluctuations inside Bob's station in a CV-QKD implementation.}
\label{fig:phase}
\end{figure}

In Fig. \ref{fig:phase}, we plot the graph showing the phase drift inside Bob's station (from point A to BS in Fig. \ref{fig:cvqkd}) being random and slow. This is experimentally measured, with the data taken from an asymmetric Mach-Zehnder interferometer, having 100 ns delay between the paths of signal and LO. This in principle mimics the phase drift inside Bob's station. We can see the change in the quadrature values, wherein the output is measured for certain intervals, as shown. One can think of this as effectively restarting the measurement afresh at the beginning of each measurement interval. We notice that the phase drift is random in nature. Additionally, it is also evident from the figure that the variation of phase drift is slow. In particular, compared to the QKD repetition rate (MHz), the frequency of phase drift is happening on the scale of milliseconds (kHz). Therefore, over a short period of time ($\le$ milliseconds), $\theta_{\text{Bob}}$ can be treated as constant for a block of quadrature measurements. In other words, in order to estimate the phase drift by Bob, the clock rate must be comparatively higher than the rate of the phase drift for our scheme to work. Bob later discloses $\theta_{\text{Bob}}$ for each block of samples to Alice. This is then equivalent to the previous case in which the frame of measurement is fixed but unknown to Eve.

\section{Discussions and conclusions \label{sec:disc}}

The GMCS CV-QKD enables Alice and Bob to share secret keys using the amplitude and phase modulated coherent states. Bob randomly switches between the quadratures to monitor the variance of them. This prevents Eve from eavesdropping as she will have to necessarily increase the variance along certain angle even for the most optimal attack. Here, we have proposed a modified scheme, wherein the measurement of a single quadrature is performed by Bob along an arbitrary angle $\theta_{\text{Bob}}$ in the phase space. Bob estimates the angle $\theta_{\text{Bob}}$ and later, during the post-processing, discloses it over the authenticated classical channel to Alice. By knowing the angle of rotation of the frame of measurement, Alice can apply an equivalent counter rotation (-$\theta_{\text{Bob}}$), and obtain a single quadrature value that coincides with that at Bob. The security of this single quadrature measurement scheme lies in the fact that Eve (as well as Alice) has no knowledge of the arbitrary angle $\theta_{\text{Bob}}$ during the quantum states transmission period. We also show that the eventual key rates with the proposed modification would yield identical results to that of GG02 protocol, specifically when $\theta_{\text{Bob}}$ is fixed. This is essentially due to the fact that when Alice's and Bob's reference frames are correlated, the maximum yield is same as GG02 protocol. However, as seen in Fig. \ref{fig:key}, a phase drift of $5^\circ$ would correspond to excess noise of $\xi=0.0107$ resulting in a reduced key. In the simplest case of intercept and resend attack \cite{bennett1992experimental}, Eve necessarily would have to squeeze along a certain angle of the intercepted signal, as she would introduce noise and increase the variance along that angle. And this will be observed by Bob as excess noise, and hence cannot evade her detection.

We consider the case of collective attack by Eve using an entangled cloner, wherein she entangles an ancillary mode to the coherent state sent by Alice. Her best strategy would be to measure her ancilla after the public announcement of the measurement basis, here the angle $\theta_{\text{Bob}}$. The announcement of $\theta_{\text{Bob}}$ by Bob is equivalent to his announcement of choice of quadrature for the measurement in GG02 protocol. Here we map our scheme to an entanglement based one, such that Alice prepares a TMSVS state and Eve entangles one of her modes of squeezed vacuum state with it to extract information. We notice that the reduction in entanglement between mode-1 and mode-2 of the TMSVS is correlated with the difference between the variances of the Alice's and Bob's quadratures, as given in Eq. (\ref{eq:sigmaprimeab}). Thus, any such attack by Eve would change the variance of mode-2, and which is quantifiable for estimating the information leakage to Eve.

Another possible attack by Eve would be having multimode entanglement with mode-2, instead of using TMSVS of Eq. (\ref{eq:evetmsvs}). In this case, Eve would necessarily use an entangled cloner of higher dimension, so that she can measure along all the possible values of $\theta_{\text{Bob}}$. This would be similar to Eve using multimode in the case of a typical GMCS QKD protocol and measure along both the quadratures \cite{zhang2024continuous}. This would enable her to get the information before Bob's announcement of the measurement quadrature ($\theta_{\text{Bob}}$ in our case). However, she will have to squeeze the state along a certain angle and the footprint of Eve's entanglement in mode-2 is inevitable, observed as the increase of variance $V(\hat{x}_{\text{B}})$ in Eq. (\ref{eq:sigmaprimeab}). Thus, any entanglement from Eve would necessarily increase the variance and this is inevitable. The primary advantage of using multimode entanglement would be to overcome the hurdle of waiting for Bob's announcement. Note that though Eve is enabled to have a cloner of higher dimension, it is still the collective attack.

The primary assumption in our scheme is that Bob needs an arbitrary angle of rotation in his frame of reference, which Eve has no knowledge of it beforehand. This is similar to the assumption of having random measurement of one of the quadrature, which Eve could not possibly influence. The proposed scheme is shown to be as feasible and secure as GMCS protocol, under the collective attack by Eve. Also, the proposed scheme is different from unidimensional CV-QKD protocols \cite{usenko2015unidimensional, usenko2018unidimensional, wang2018security, bai2020unidimensional}, in that, the state preparation in these protocols involve modulation of a single quadrature. Alice modulates only one quadrature in these protocols, but Bob measures both the quadratures to verify the uncertainty between them. In contrast, the Alice setup in our case remains to be identical to the GG02 protocol, and so is the proposed scheme more closer to it. 

We note here that the single quadrature measurement scheme in this work is close to that in Ref. \cite{liu2018continuous}. However, the basis of choice by Bob is continuously drifting for every hundreds of measurements in the former. In particular, Bob chooses $\theta_{\text{Bob}} \in [0,2\pi]$ such that it is continuously changing. Whereas, in the current work, it can be kept constant for the whole run. In the GMCS protocol with homodyne measurement, for the case when Bob always measures in single quadrature and later announces the choice, Eve's success probability to hide her attack is 50\%. One can think of a similar attack in the present case as well, wherein Eve too randomly chooses her rotated frame of measurement to be an arbitrary angle. But this is a non-optimal attack, as Eve's success rate reduces with the increase in the resolution of the phase space of Bob's apparatus. Ideally, the probability to detect Eve's presence is
\begin{equation}
P_{\text{Bob}} = \frac{1}{N_{\{i\}}} \sum_{i} p_i,
\end{equation}
where $p_i$ is the respective probability of the phase space resolutions and $N_{\{i\}}$ is the resolution dependent normalization factor. We note that Eve is unaware of Bob's angle of rotation in frame of reference (and the phase drift), until Bob's public announcement of the same. Let the arbitrary angle of measurement by Eve be $\theta_{\text{E}}$. Thus, only at the angle $\theta_{\text{Bob}} = \theta_{\text{E}}$, $p_j = 0$. As $\abs{\theta_{\text{Bob}} - \theta_{\text{E}}}$ increases, Eve's information decreases and Bob's probability to detect her increases. 

Note that the above mentioned attack is neither optimal nor realistic. Eve could further reduce $P_{\text{Bob}}$ if she performs this non-optimal attack probabilistically. In particular, only on a fraction $f$ of the total transmissions, she could do this attack along a random angle. The primary advantage of this attack is that the footprint that she leaves is a fraction of the earlier case. That is, the probability of detection by Bob in this case would be $P^\prime_{\text{Bob}} = f\cdot P_{\text{Bob}}$. However, the information that she gets is also reduced by the same factor, as ideally the success probability is equal for all the pulses. Below we quantify the normalized information that Eve would get and the corresponding probability of Bob finding about Eve's presence, for $f=1$.

\begin{figure}[h]
\includegraphics[width=\linewidth]{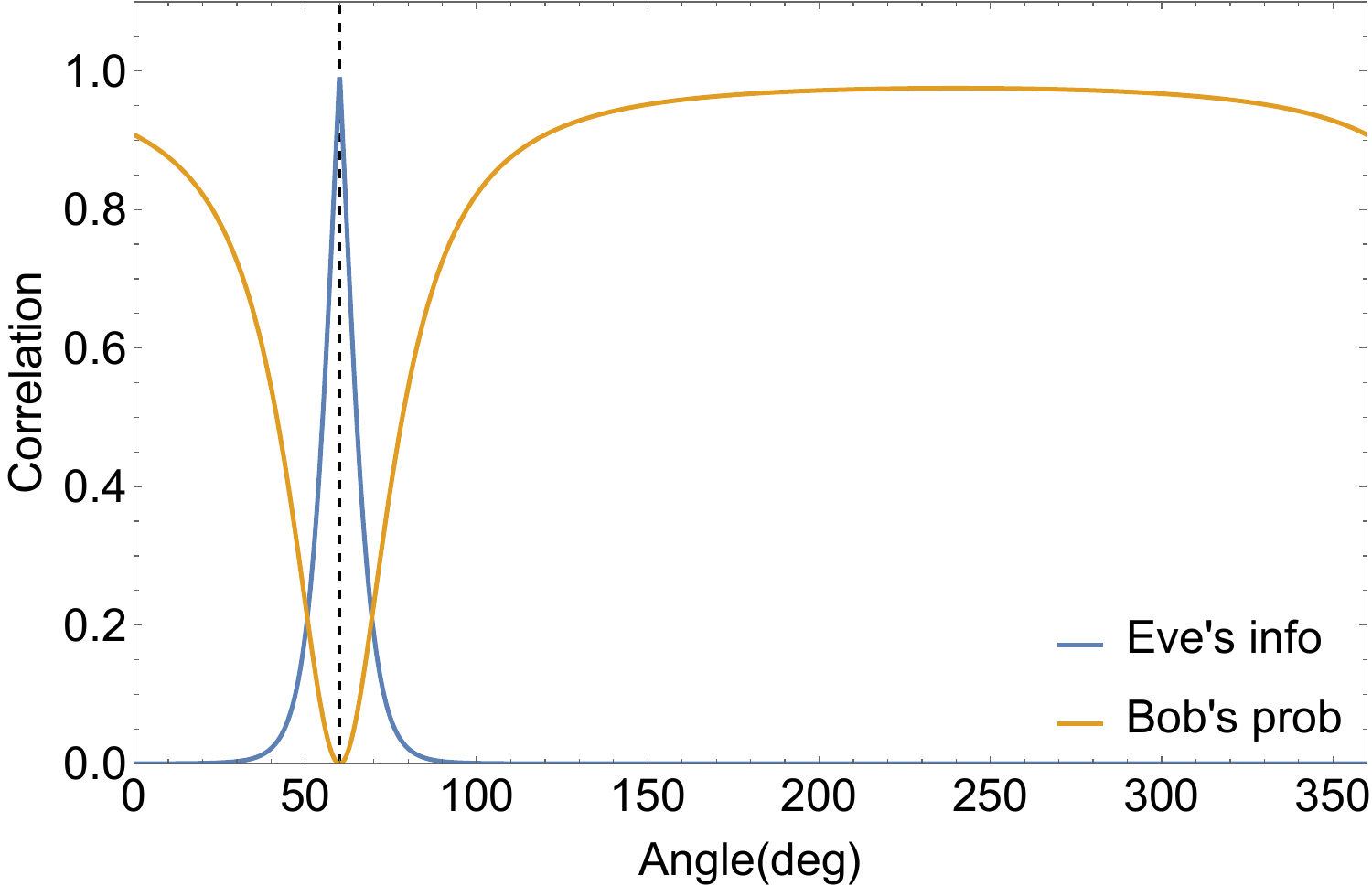}
\caption{The Blue line indicates the normalized information of Eve w.r.t the choice of her measurement angle. The Orange line indicates the normalized probability of Bob finding about Eve's presence. Here the normalization is based on the phase space splicing by Bob, with the angle of measurement is taken to be $60^\circ$.}
\label{fig:angle}
\end{figure}

In the above case, let the resolution in the phase space be $\frac{1}{2\pi}$ for both Bob (and Eve). We find that the corresponding coefficient of the quadrature value for Eve is a function of $\cos{(\theta_{\text{E}}\pm\theta_{\text{Bob}})}$, with $\pm$ indicating the relative angle. Thus, the correlation between information leakage due to Eve's measurement and Bob's probability of finding squeezing by Eve is given in blue in Fig. \ref{fig:angle}. Here the normalization is due to the resolution in phase space. Let us consider the usual squeezing parameter quantifying the reduced variance with $V = 10^{\frac{-V_{\text{dB}}}{10}} = e^{-2r}$. Also, $\chi=re^{i\theta_{\text{Bob}}}$ is the squeezing parameter along the direction of $\theta_{\text{Bob}}$. Thus, if we consider a total squeezing of $V_{\text{dB}}=$10 dB by Eve, which corresponds to the squeezing parameter being $\chi=1.15e^{i\theta_{\text{E}}}$. Thus, we find the probability of detecting Eve's presence to be,

\begin{equation}
P_{\text{E}}=\frac{1}{2\pi}\bigg[\frac{\mathcal{V}_{\text{B}}}{\abs{\cos(\pi+(\theta_{\text{E}} \pm \theta_{\text{Bob}})/2)}}\bigg],
\end{equation}
where $\mathcal{V}_{\text{B}}$ is the limiting factor for shot-noise measurement of Bob's detector. This is given in orange in Fig. \ref{fig:angle}. Thus, the success probability of Eve would be much lower in this case, and is quantified by the resolution of phase space and the respective weightages. Here we note that as the resolution of the phase space becomes finer, Eve's probability of concealing her presence decreases.

The current work limits its arguments and results to the Gaussian distribution-based protocol. The necessity of switching between the two quadratures is relaxed in a homodyne detection scheme and the protocol is shown to be identical to GG02 protocol \cite{grosshans2002continuous}. The announcement of the angle of rotation by Bob is similar to that of announcement of chosen quadrature in the latter protocol. This works for the LLO based systems as well as the TLO based system used in this analysis. Conventionally, for LLO scheme, a homodyne detection setting is used for signal and heterodyne for reference pulses \cite{soh2015self}. In TLO scheme, the phase drift is estimated using the pattern signals which are comparatively at higher intensity than the QKD signals and are sent with each block of QKD signals \cite{cosijns2018advanced, tang2021towards, tang2022experimental}. The homodyne detection of signal in LLO scheme is particularly for reaching maximum transmission distance, however it still involves switching between the two quadratures. Thus, our scheme avoids the use of phase modulator for homodyne detection of the signal, in GG02-like protocols, regardless if a TLO or LLO is implemented. Thus the proposed modification simplifies the protocol for CV-QKD systems that have homodyne detection setting as it overcomes the necessity of a randomly varying phase modulator at Bob's station. This is highly beneficial for practical implementation, especially in free-space CV-QKD with Gaussian modulation protocol. 

\begin{acknowledgments}
The authors thank Frédéric Grosshans for the useful discussions. V.N.R, T.S and R.K acknowledge the funding support from EPSRC Quantum Communications Hub (Grant number EP/T001011/1). E.T.H.M thanks the School of Physics, Engineering and Technology, University of York for PhD funding.
\end{acknowledgments}

\bibliography{References}

\appendix

\section{Covariance matrix \label{app:calc}}

To find the covariance matrix of a two-mode squeezed vacuum states (TMSVS), we follow the standard formalism. In particular, each element of the covariance matrix at different stages are found as the follows. The diagonal ones are found using $\langle \hat{\mathcal{Q}}_n^2 \rangle - \langle \hat{\mathcal{Q}}_n \rangle^2$, where $\mathcal{Q}$ is the quadrature chosen for measurement in mode $n$. The off diagonal elements are similarly found, except it is the average of the expectation values of the product of the modes as $1/2(\langle \hat{\mathcal{Q}_n}\hat{\mathcal{Q}_m} \rangle + \langle \hat{\mathcal{Q}_m}\hat{\mathcal{Q}}_n \rangle)$.

We find that the joint matrix of system ABE after Eve's entanglement with mode-2 to be,
\begin{align}
\Sigma^\prime = 
\begin{pmatrix}
V\identity_2 & \mathcal{X}_{+}P & \mathcal{X}_{-}P & 0 \\
\mathcal{X}_{+}P^{\text{T}} & \mathcal{Z}_1 & \mathcal{Z} & \mathcal{Y}_{-}Q \\
\mathcal{X}_{-}P^{\text{T}} & \mathcal{Z} & \mathcal{Z}_2 & \mathcal{Y}_{+}Q \\
0 & \mathcal{Y}_{-}Q^{\text{T}} & \mathcal{Y}_{+}Q^{\text{T}} & W\identity_2+\frac{(W+1)}{2}S
\end{pmatrix},
\label{eq:sigmaprime}
\end{align}

where,
\begin{align*}
\mathcal{X}_{+} &= \sqrt{T(V^2-1)}, \\
\mathcal{X}_{-} &= -\sqrt{(1-T)(V^2-1)}, \\
\mathcal{Y}_{-} &= \sqrt{(1-T)(W^2-1)}, \\
\mathcal{Y}_{+} &= \sqrt{T(W^2-1)}, \\
\mathcal{Z} &= \sqrt{T(1-T)}\big[W\identity_2 - (V\identity_2 +R)\big], \\
\mathcal{Z}_{1} &= T(V\identity_2 +R) + (1-T)(W\identity_2), \\
\mathcal{Z}_{2} &= (1-T)(V\identity_2 +R) + T(W\identity_2).
\end{align*}

\section{Quantifying Eve's information \label{app:mut}}

Given Alice prepares TMSVS and keeps one mode to herself while sending another to Bob, we find that the commutation relation between them would be,
\begin{equation}
[\hat{x}_j,\hat{x}_k] = 2i\Omega_{jk},
\end{equation}
where $\Omega = \oplus \{\{0,1\},\{-1,0\}\}$ is a 4x4 standard anti-symmetric matrix, and $\hat{x} = (\hat{q_{\text{A}}},\hat{p_{\text{A}}},\hat{q_{\text{B}}},\hat{p_{\text{B}}})^{\text{T}}$ is the displacement vector. When the covariance matrix is of the form
\begin{align}
V=
\begin{pmatrix}
A & C \\
C^{\text{T}} & B
\end{pmatrix},
\label{eq:cova}
\end{align}
with $A=A^{\text{T}}$ and $B=B^{\text{T}}$, the symplectic eigenvalues are given by, 
\begin{equation}
\nu_{\pm} = \sqrt{\frac{\mu \pm \sqrt{\mu^2 - 4~(\text{det} V)}}{2}}
\label{eq:nupm}
\end{equation}
where $\mu:= (\text{det} A) + (\text{det} B) + 2(\text{det} C)$ and $\text{det}(\cdot)$ is the determinant. Thus, the von Newmann entropy of $\Sigma^\prime_{\text{AB}}$ is given by $S_{\text{AB}} = g(\nu_{+}) + g(\nu_{-})$, where 
$$g(\cdot) = \bigg(\frac{\cdot + 1}{2}\bigg) \log_2\bigg(\frac{\cdot + 1}{2}\bigg) - \bigg(\frac{\cdot - 1}{2}\bigg) \log_2\bigg(\frac{\cdot - 1}{2}\bigg).$$

Similarly, we find $S_{\text{A}\mid \text{B}}$ by the symplectic eigenvalues of the covariance matrix $\Sigma^\prime_{\text{A}\mid \text{B}}$, taking the partial measurements on the covariance matrix of Eq. (\ref{eq:cova}). In particular, for a homodyne detection at Bob's station, we define the partial measurement of mode-2 transforming mode-1 as,
\begin{equation}
\Sigma^\prime_{\text{A}\mid \text{B}} = A - C(\Pi_x ~B~ \Pi_x)^{-1}C^\text{T},
\label{eq:nu}
\end{equation}
where $\Pi_x = \{\{\cos{\theta},0\},\{0,\sin{\theta}\}\}$ being the measurement setting of Bob and $(\cdot)^{-1}$ indicating the pseudoinverse. Thus, we find the symplectic eigenvalue $\nu$ of $\Sigma^\prime_{\text{A}\mid \text{B}}$, and $S_{\text{A}\mid \text{B}} = g(\nu)$. 

\section{Generalized case \label{app:gen}}

All the earlier cases involves Alice measuring in standard quadratures and Bob in his rotated frame of measurement. However, in principle, Alice too might measure in an arbitrary rotated angle and later establish the correlation of the angle with Bob. This would be a general case, wherein the measurement of TMSVS is $\theta_{\text{A}}$ for mode-1 and $\theta_{\text{B}}$ for mode-2. The covariance matrix for such a system is
\begin{align}
\Sigma_{\text{AB}} =
\begin{pmatrix}
&V\identity_2+\frac{(V+1)}{2}R_1^\prime & \sqrt{V^2-1}P_1 \\
&\sqrt{V^2-1}P_1^{\text{T}} &V\identity_2+\frac{(V+1)}{2}R_1
\end{pmatrix},
\end{align}
where $R_1^\prime=\{\{0,\sin{\theta_{\text{A}}}\},\{\sin{\theta_{\text{A}}},0\}\}$, $P_1=\{\{CS,-\sin{\theta_{\text{A}}}\},\{-\sin{\theta_{\text{B}}},-1\}\}$ and $R_1=\{\{0,\sin{\theta_{\text{B}}}\},\{\sin{\theta_{\text{B}}},0\}\}$, with $CS = \cos{\theta_{\text{A}}}\cos{\theta_{\text{B}}} - \sin{\theta_{\text{A}}}\sin{\theta_{\text{B}}}$. The consequent joint system of ABE would be similar to Eq. (\ref{eq:sigmaprime}), but with different variances. However, after the announcement of $\theta_{\text{B}}$ and Alice rotating her frame to match, the respective mutual information and leaked information would be identical to the earlier case.

\end{document}